\documentclass[preprint2]{aastex}

\slugcomment{Version: 7 Sep 2001}

\shorttitle{X-ray Emission from Radio Jets}
\shortauthors{Harris \& Krawczynski}

\begin{document}


\title{X-ray Emission Processes in Radio Jets}


\author{D. E. Harris,}
\affil{Smithsonian Astrophysical Observatory, 60 Garden St. Cambridge,
MA 02138}
\email{harris@cfa.harvard.edu}

\and

\author{H. Krawczynski,}
\affil{Yale University, New Haven, CT}
\email{krawcz@astro.yale.edu}





\begin{abstract}

The emission processes responsible for the observed X-rays from radio
jets are commonly believed to be non-thermal, but in any particular
case, it is unclear if synchrotron emission or one or more varieties
of inverse Compton emission predominates.  We present a formulation of
inverse Compton emission from a relativistically moving jet
(``IC/beaming'') which relies on radio emitting synchrotron sources
for which the energy densities in particles and fields are comparable.
We include the non-isotropic nature of inverse Compton scattering of
the relativistic electrons on photons of the cosmic microwave
background (CMB) and provide beaming parameters for a number of jets.
A list of X-ray emitting jets is given and the jets are classified on
the basis of their morphology and spectral energy distribution to
determine their likely emission process.  We conclude that these jets
have significant bulk relativistic velocities on kpc scales; that
higher redshift sources require less beaming because the energy
density of the CMB is significantly greater than locally; and that for
some nearby sources, synchrotron X-ray emission predominates because
the jet makes a large angle to the line of sight.

\end{abstract}


\keywords{galaxies: jets---magnetic fields---radiation mechanisms: non-thermal}


\section{Introduction}\label{sec:intro}

X-ray emission associated with radio jets in extragalactic sources now
comes in a large variety of diverse characteristics.  When this sort
of emission was first isolated (e.g. the radio galaxy M87 with the
EINSTEIN OBSERVATORY, Schreier, Gorenstein, \& Feigelson 1982) and
there were only a few examples, it was tempting to work on the
assumption that the emission process was definable and would apply to
all examples.  This notion persisted into the ROSAT era until a
convincing case was made that the terminal hotspots of Cygnus A
represented synchrotron self-Compton (SSC) emission (Harris, Carilli,
\& Perley, 1994).

With the advent of the CHANDRA OBSERVATORY, the number of sources has
almost tripled (from 7 to at least 19) yet there remain substantial
problems in determining the emission process responsible for the X-rays in
some sources.  Although broken power laws connecting the radio,
optical, and X-ray data are still viable spectral models for a few
sources, a number of other sources are observed to have such a low
flux density in the optical that the indicated cutoff in the spectrum
would preclude a simple connection to the X-ray data.

The introduction of the 'beaming model' (Tavecchio et al. 2000,
Celotti et al. 2001) in the particular case of PKS0637 was presented
as an escape from this dilemma.  In this model, the enhancement of the
X-ray emission relative to the synchrotron (radio/optical) is
explained by hypothesizing a bulk velocity of the jet fluid which is
relativistic even on large (kpc) scales.  If this were the case, in
the frame of the jet the effective photon energy density of the cosmic
microwave background (CMB) would be augmented by the square of the
jet's Lorentz factor, $\Gamma$, and it was demonstrated that quite
reasonable combinations of $\Gamma$ and the beaming factor, $\delta$
could be invoked to explain the observed intensities.

In this paper we review the common emission processes
(section~\ref{sec:emission}); present a current list of jet sources
and suggest a classification scheme (section~\ref{sec:classify});
describe a formulation of the beaming model which relies on an
evaluation of the magnetic field strength from the equipartition field
and includes the anisotropic nature of the IC scattering
(section~\ref{sec:beam}); and examine the conflicting evidence and
ramifications for the synchrotron and inverse-Compton (IC) models
(section~\ref{sec:conflict}).  The appendix contains the details of
our beaming formulation.

For numerical results we use cgs units unless stated otherwise and
assume H$_0$=50~km~s$^{-1}$~Mpc$^{-1}$; and q$_0$=0.  We follow the
convention that the spectral index of a power law is defined by flux
density, S$_{\nu}~=~k~\nu^{-\alpha}$.

\section{Overview of emission processes}\label{sec:emission}

\subsection{Thermal bremsstrahlung}

For many sources which display convincing evidence that the X-ray and
radio emission originate in the same volume, (e.g. hotspot B of
3C390.3, Harris Leighly \& Leahy 1998) it has been argued that thermal
bremsstrahlung does not provide a satisfactory model for the X-ray
emission because the required amount of hot gas is large
($\approx~10^{10}M_{\odot}$), over-pressured, far from the parent
galaxy, and the predicted Faraday rotation and depolarization are not
observed.  Recent Chandra results (e.g. 3C 273, Marshall et al. 2001a)
confirm that the X-ray emission from jet features has no line emission
and is best characterized by a power law.

\subsection{Synchrotron emission}\label{sec:sync}

Synchrotron emission has been considered the `process of choice' for
X-rays from knots in radio jets mainly because the optical
polarization observed in sources such as M87 is convincing evidence
that the optical emission as well as the radio emission comes from the
synchrotron process.  Demonstrations that the X-ray intensity was
consistent either with a single power law extrapolation from radio and
optical bands (e.g. hotspot B of 3C~390.3, Harris, Leighly, \& Leahy
1998) or with a broken power law (e.g. knot A in M87, Biretta, Stern,
\& Harris 1991) were taken as circumstantial evidence that the X-rays
were also generated by synchrotron emission.  Required for this model
is the presence of electrons with Lorentz factor $\gamma > 10^7$ (cf.
values of 10$^5$ for optical emission).  In the typical equipartition
fields of B $\approx 10^{-4}$G, the radiation half lives $\tau_o$ of
the X-ray emitting electrons would be of order 10 years (however, see
Aharonian (2001) for problems associated with very fast cooling
times).

There are, however, a number of sources for which the optical flux
densities or limits preclude a simple construction of a broken power
law.  In general, assuming the usual shock acceleration processes and
dominance of losses to the relativistic electrons that go with the
energy squared, we expect that both the electron distribution and the
resulting synchrotron emission spectrum will be concave downward when
displayed on the usual log~S$_{\nu}$ vs. log~$\nu$ plot.  For this reason,
X-ray intensities that lie well above the extrapolation of the
radio/optical synchrotron spectrum are taken to be strong evidence
against the 'simple' synchrotron model.

The emission of most of these sources however can still be explained
with inhomogeneous synchrotron models.  Spatially separated emission
components which cannot be resolved with the current X-ray detectors
would be the consequence of diffusive shock acceleration with a time
dependent high energy cut-off of accelerated particles.  While the low
energy radio and optical emission is dominated by radiatively aged
particle populations further downstream, the X-ray emission is only
emitted by recently accelerated particles.  In this scenario the
projected extension of the X-ray emission region is either given by
the projected size of the accelerating shock, or by the distance the
downstream plasma travels before the highest energy electrons cool
away.  For 3C~120 such an interpretation encounters the difficulty
that the optical to X-ray energy spectrum is as hard as $\alpha_{\rm
OX}<$0.35 which indicates a spectrum of accelerated particles with a
spectral index, $p^<_\sim 1.7$ (even in the optimistic case that the
X-ray electrons did not cool substantially); harder than the canonical
values of 2 expected for acceleration at strong non-relativistic
shocks.

Another class of models involves high energy protons.  These models
and their difficulties are described in Mannheim, Krulls, \& Biermann
(1991) and Aharonian (2001).

\subsection{Inverse Compton emission}

Inverse Compton emission has a distinct advantage over X-ray
synchrotron emission in that extremely high energies for the
emitting electrons are not required.  What is required are both enough
electrons and enough energy density in photons of the proper energy to
produce the desired scattered photons:
$\nu_{out}~\approx~\nu_{in}~\times~\gamma^2$.  

\subsubsection{Synchrotron self Compton emission}\label{sec:ssc}

Unlike synchrotron emission for which the radio data often provide no
clear indication about the higher energy parts of the electron
spectrum, accurate predictions of source intensity can be calculated
for SSC emission.  Given a reasonable estimate for the emitting
volume, it is possible to calculate the photon energy density from the
synchrotron spectrum and then determine what B field is required to
have the right number of electrons to give the observed (radio)
synchrotron and the observed (X-ray) IC emissions.

Although SSC emission is mandatory for all synchrotron sources, it is
usually the case that the photon energy density ($u_{\nu}$) is
significantly smaller (i.e. by factors of 100 or more) than the energy
density of the magnetic field ($u_B$) and thus the major energy loss
for all electrons is via synchrotron emission (and the SSC component
is too weak to observe or distinguish from other emissions).

There are however, four FRII radio galaxies with convincing SSC X-ray
emission from their (terminal) hotspots: Cyg A (Harris et al. 1994;
Wilson,, Young, and Shopbell 2001a); 3C~295 (Harris et al. 2000);
3C~123 (Hardcastle, Birkinshaw, \& Worrall 2001a), and 3C 263
(Hardcastle et al. in preparation).  For all of these sources,
predicted SSC emission was calculated from the radio data, proposals
were written, and the hotspots were detected as predicted.  In all
cases, the average magnetic field strengths derived from the SSC
equations are consistent with equipartition fields for the case of
little or no contribution to the particle energy density from
relativistic protons.  The hotspots of these sources are, so far, the
only resolved structures for which convincing SSC models have been
published.

\subsubsection{IC emission from the microwave background}

Except for high redshift sources, IC emission from the relativistic
electrons scattering off the CMB is incapable of producing the
observed intensity of X-rays unless the beaming model is invoked with
relativistic values of the bulk jet velocity (see
section~\ref{sec:beam}).  The beaming model is attractive because most
or all of the X-ray emitting jets are one sided and the model permits
the relative increase of the effective photon energy density to that
in the magnetic field, thereby boosting the emission in the IC channel
relative to that in the synchrotron channel.

\subsubsection{IC emission from other photons}

External photons have been employed successfully in explaining X-ray
emission from some radio lobes (Brunetti et al. 2001).  However, it is
unlikely to apply to knots far removed from the hypothesized highly
luminous quasar-like core and Brunetti's model provides IC enhancement
on the receding side of the source, not the approaching side as
observed in our sample.  

Celotti et al. (2001) discuss the possibility that emission from a
high velocity `spine' of a jet may serve as seed photons for IC
emission from a lower velocity sheath.  Generally, a wide latitude of
inventiveness is allowed for IC emission since the dominant
contribution to the photon energy density may be anisotropic and hence
not directly observable.

\section{Classification of sources}\label{sec:classify}

The contents of this section are `subject to change'
since much of our information is incomplete.  The reason for including
this source list is to provide an overview.  Some aspects of the
classification are subjective and we anticipate changes as new
data become available.  We maintain current information at
\url{http://hea-www.harvard.edu/XJET/}.

Table~\ref{tab:sources} contains a list of bona fide X-ray jet sources
known to us as of 2001 March.  In the last column is the
classification which is described here.


\subsection{SSC}

This category is perhaps more secure than the others because the data
confirm the predictions rather well.  All 4 SSC sources
(sec.~\ref{sec:ssc}) are FRII radio galaxies and their terminal
hotspots rather than brightness enhancements in the jets are the SSC
emitters.  For all of these, the magnetic field strengths in the
hotspots are in the range 100 to 400 $\mu$G, values consistent with
the equipartition field for the case of little or no contribution to
the particle energy density from protons.  Synchrotron luminosities
and lifetimes are unaffected by the IC losses and the electrons
responsible for the observed X-rays are those emitting synchrotron
emission in the normal radio band.  Alternative models for these
hotspots are the generation of high energy electrons via the PIC
(Proton Induced Cascade, Mannheim et al. 1991) or proton
synchrotron (Aharonian 2001) processes, both of which would require much
stronger magnetic fields ($B\,\gg\,500\mu$G).

\subsection{Synchrotron}

Mindful of ongoing debate about the evidence for spectral cutoffs from
optical data (sec.~\ref{sec:sync}), we classify M87 knots D, A, and B as
synchrotron emitters (Biretta, Stern, and Harris, 1991; Marshall et
al., 2001b).  Additional sources in this category are 3C 390.3 hotspot
B (Harris et al. 1998), knots A1 and B1 in the 3C 273 jet (Marshall et
al. 2001a), and 3C 66B (Hardcastle, Birkinshaw, \& Worrall, 2001b)
although for some of these, there is the problem that a cooling break
in the emission spectrum has not been isolated.

The extension of the synchrotron spectrum from the optical to the
X-ray has only a small effect on the total energy and on
the calculation of the equipartition magnetic field.  However, this
model requires the extension of the electron spectrum from
$\gamma~\approx~10^5$~to~$\geq~10^7$.  The primary observable
consequence is that the electrons responsible for the X-rays have
lifetimes of order 10 to 100 years.

If a somewhat 'ad hoc' additional spectral component is allowed
(sec.~\ref{sec:distinct}; i.e. a high energy component of the electron
energy distribution with a flatter spectrum than that observed in the
radio/optical domain), then synchrotron emission models can be devised
for sources such as Pictor~A (Wilson, Young, \& Shopbell 2001b,
section 4.2.2.2), 3C 273 knot D, PKS 0637, and 3C~120 (Harris et
al. 1999; where a very hard spectral index of accelerated particles of
$p$=1.7 would be required).  The SSC sources (sec.~\ref{sec:ssc})
could also be accomodated by synchrotron models if additional spectral
components are allowed.

\subsection{Bulk relativistic velocities}

In sections~\ref{sec:beam} and \ref{sec:conflict} we discuss many of
the details and ramifications of this model.  While there is little
debate concerning the utility of the beaming model in its ability to
explain a larger ratio of X-ray to synchrotron emission than other
models, there is no completely convincing evidence that kpc-scale jets
actually have bulk velocities close to c.

Beaming models have been presented for PKS0637 (Celotti et al, 2001;
Tavecchio et al. 2000) and for 3C 273 (Sambruna et al. 2001).  For the
knots in these jets, viewing angles are generally required to be less
than 10$^{\circ}$, $\Gamma$ values range from 5 to 20, and
equipartition fields are significantly smaller than for unbeamed
synchrotron emission.  A very significant difference between
beaming and synchrotron models is that the beaming model posits that
the electrons responsible for the observed X-rays are at the very
bottom of the electron distribution with $\gamma$ in the range 20-150.
This implies that no variability is expected in the X-ray intensity.
Because there is a significant difference between the electron
energies producing the X-rays and those that produce radio synchrotron
emission observable from the Earth with sufficient resolution to
measure accurate spectral indices, there is a large uncertainty in the
expected spectral shape of the electron distribution at low energies,
and consequently, in the predicted spectral shape and the intensity of
the X-ray emission.

\subsection{Discrepant radio/X-ray morphologies}

Although most of the sources in Table~\ref{tab:sources} display
spatial coincidence between the radio, X-ray, and optical
morphologies, there are a few sources where this is not the case.  Cen
A, 3C 66B, and PKS 1127 are the three examples of this behavior in our
collection.  We do not believe that this effect results from limited
angular resolution since it occurs in both the nearest (Cen A) and
furthest (PKS1127) source.  In the case of Cen A, some offsets
between the peak brightness of radio and X-ray are of order
2$^{\prime\prime}$ (34pc) while other features are coincident.  For
PKS1127, an offset of 0.8$^{\prime\prime}$ observed for knot B
corresponds to 9 kpc.

For these jets, it is difficult to measure radio and x-ray intensities
for well defined volumes.  It is also the case that the PKS1127 jet is
very weak in the radio (the source is a VLA calibrator) and the
published radio data for Cen A do not have the combination of high
resolution and s/n which are necessary for reliable inter-band
comparisons.  For 3C 66B and PKS1127, the peak X-ray brightness occurs
upstream of the peak radio brightness. To obtain significant X-ray
emission with little or no radio emission would require extreme
beaming parameters or a distinct low energy electron population
(sec.~\ref{sec:distinct}) for the beaming model or a flat high energy
component of the electron spectrum for synchrotron models.

\subsection{Not yet classified}

Since data have not yet been published, we have not classified 3C 15,
3C 31, PKS0521, and NGC4261 in Table~\ref{tab:sources}.

\section{Determination of jet parameters within the beaming model}\label{sec:beam}

The beaming model introduced by Tavecchio et al. (2000) and Celotti et
al. (2001) is appealing because it offers a method of increasing the
IC emission relative to the synchrotron emission and because it was
already clear from EINSTEIN and ROSAT results on M87, 3C 120, and 3C
390.3 that all known X-ray emitting jet features were one sided.  Most
of the new examples from Chandra detections maintain the one sided
nature of X-ray emission, a natural consequence of the beaming model.

The following formulation is based on the assumption of equipartition
in the synchrotron source in order to evaluate the magnetic field
strength and accounts for the anisotropic nature of the IC emission in
the jet frame.  The details are given in the Appendix.
 
We define the constraints on the bulk relativistic flow velocity of
the jet fluid, $\Gamma$; the viewing angle of the line of sight of the
observer with respect to the jet velocity vector, $\theta_{\rm j}$;
and the relativistic Doppler factor, $\delta$.  Once these constraints
have been evaluated, we examine a number of source parameters:

\begin{itemize}

\item{the segments of the electron energy spectrum responsible for
both the synchrotron and IC emissions to make sure we end up with a
self-consistent source model.}

\item{the halflives of the various energy electrons.}

\item{the basic parameters of the synchrotron source in the jet
frame.}

\end{itemize}

\subsection{The case for equipartition}

Over the years since its introduction (Burbidge, 1956), the concept of
'minimum energy' (or 'equipartition') has generated considerable
debate and some confusion.  Our view is that although it was
originally borne of twin desperations (the staggering amount of
non-thermal energy required for large synchrotron radio sources and
the realization that it was almost impossible to disentangle the two
primary parameters of synchrotron emission), a reasonable case can be
advanced that most extragalactic synchrotron sources are not more than
a factor of a few from having their average magnetic field energy
density equal to the average energy density in relativistic particles:
u(p)~$\approx$~u(B).

Obviously we are unconcerned with the condition u(p) $<<$ u(B); it is
the converse that has generated some interest as a means of increasing
the relative emission in the IC channel compared to that in the
synchrotron channel.  For radio structures far from their parent
galaxy it seems that the only mechanism which serves to maintain the
integrity of a radio lobe is the magnetic field, acting as a boundary
between the non-thermal plasma and the external thermal plasma.
Direct evidence that some synchrotron emitting plasmas maintain their
identity is provided both by the detection of cavities in cluster gas
(Carilli, Perley, \& Harris 1994) and by the observation that the B
field direction is normally tangent to the edges of radio lobes (and
not tangled as expected if turbulent mixing were occurring).  If
the particle energy density were much greater than u(B) it is unlikely
that the field would confine the particles and radio sources would be
transient phenomena without well defined boundaries.

Evidence for equipartition has come from the analyses of X-ray
emission from terminal hotspots.  In the study of Cygnus A hotspots,
Harris et al. (1994) argued that if the observed X-ray emission were
to be SSC emission, then the average magnetic field strengths
(B$_{\rm IC}$) would be very close to the equipartition fields estimated
from the conditions: filling factor = 1 and no contribution to the
particle energy density from relativistic protons.  Since we have a
fairly accurate estimate of the synchrotron photon energy density, we
know that if there were more electrons than those corresponding to
B$_{eq}$, we would over-produce the X-ray emission.  Thus the derived
value of B$_{ic}$ is a firm lower limit and the only method of driving
the hotspots out of equipartition with u(p) $>>$ u(B) would be to add
non-radiating particles.  The three recent additions to the collection
of resolved SSC emitters strengthen this evidence.

A number of additional arguments have been presented in favor of
synchrotron sources being close to equipartition.  Readhead (1994)
used the observed distribution of brightness temperatures in support
of equipartition but did not deal with the physical mechanism
responsible.  
Bodo, Ghisellini \& Trussoni (1992) refined earlier work by
Syrovatskii (1969) and Singal (1986) on the effects of spiraling
electrons and their associated diamagnetic moments and stated ``The counter
field produced by a cloud of relativistic electrons lead to an
equilibrium between the magnetic field in the cloud and the energy
density of radiating electrons.''

Although it is possible to imagine synchrotron sources shortly after
the introduction of a large amount of energy in particles, it is
beyond the scope of this paper to consider the time scale required to
return to equipartition.  Since we are dealing with large ($>$kpc)
scale structures, we believe that most of the emitting regions
considered in this paper will be close to equipartition.  However,
questions are often raised such as ``What is the extent of departure
from equipartition for some given condition?''  We note from
equation~\ref{eq:resulta} that a given change in $\Gamma$ (which
involves the same change in $\delta$) will require a change in B(1) by
this factor squared, and since u(B) goes as B$^2$, the change to u(B)
goes as the fourth power.  As an example, consider the beaming required for
3C120: $\Gamma$=39 (Table~\ref{tab:results}).  How far from
equipartition would we have to go to get $\Gamma$=10?  The answer is of
order 230 times less energy in the magnetic field than in particles.
Although this calculation is approximate, the formulation presented in
the appendices relies on the equipartition assumption only to evaluate
the average field strength, and if some other estimate is available,
it can be used in place of B$_{\rm eq}'$ in eq.~\ref{eq:result}.

\subsection{Outline of the method}

We first take the synchrotron spectrum from the observer's frame back
to the jet frame.  Since the beaming factor is unknown, this process
is done (in principle) for trial values of $\delta$.  In the jet
frame, we apply the usual synchrotron formulae (e.g. Pacholczyk, 1970)
to solve for the equipartition field strength (eq.~\ref{eq:beq}).
Actually, the value of the average magnetic field for equipartition is
inversely proportional to $\delta$, so in practice, we need only
calculate $B_{\rm eq}'$ for $\delta$~=~1.

\begin{figure}
\plotone{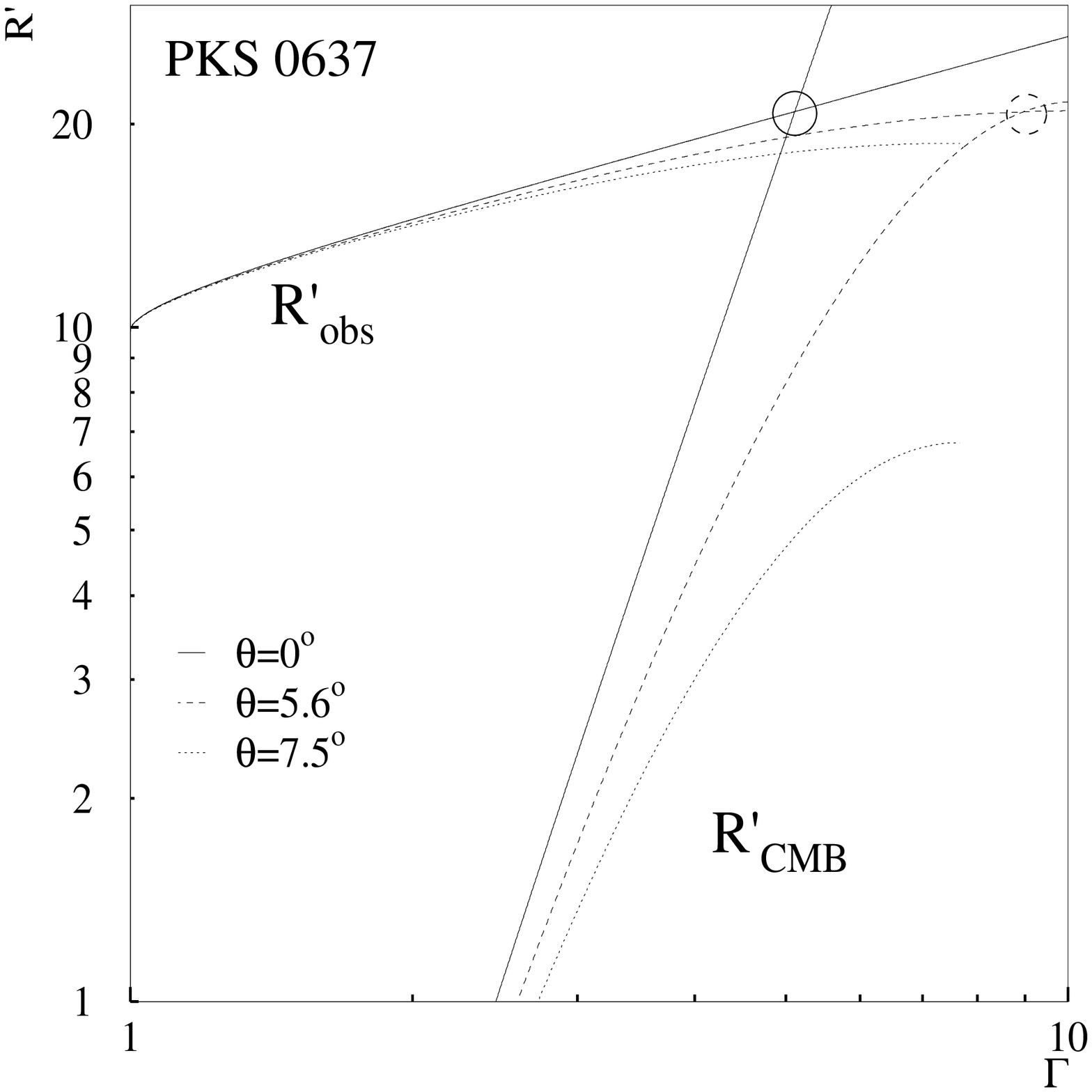}
\caption{An illustration of the solutions for
equations~\ref{eq:robs} and \ref{eq:rexp}.  The vertical axis is the
ratio of IC to synchrotron losses in the jet frame and the horizontal
axis is the Lorentz factor for the bulk velocity of the jet fluid.
For PKS0637, the range of solutions run from $\theta$~=0 to
5.6$^{\circ}$.  For larger values of $\theta$, there is no solution.  
\label{fig:method}}
\end{figure}

We then derive the expression for the IC emission (eq.~\ref{eq:power})
using the effective energy density of the CMB as seen by the jet fluid
(eq.~\ref{eq:unu}).  We evaluate the quantity R (the ratio of IC to
synchrotron luminosities) in two ways.  First, we compute R from the
integrated IC energy flux to the integrated synchrotron energy flux
(eq.~\ref{eq:robs}).  This is done by converting observed synchrotron
frequencies to electron energies (which involves the magnetic field
strength) and integrating the IC emission from the same segment of the
electron spectrum, taking into account the anisotropic radiation
pattern of the IC component.  Second, we compute the expected value of
R from the ratio of the energy densities of the CMB to that in the
magnetic field (both in the jet frame) (eq.~\ref{eq:rexp}).  Equating
these two expressions, we are able to solve for the beaming parameters
which satisfy the equality (eq.~\ref{eq:result}).  Since the jet
parameters enter the R equations in rather complex ways because of the
anisotropic nature of the IC emission, a numerical method is used.
This is demonstrated in fig.~\ref{fig:method} and we
obtain a result consistent with that of Celotti et al. (2001) for
PKS0637.

We use the following notation: we prime all quantities in the jet
frame and, in cases where ambiguities could arise, we characterize jet
parameters by the subscript ``j'' and electron parameters by the
subscript ``e''.


\subsection{Assumptions}

\begin{itemize}
 
\item{We assume that all the energy density of the CMB occurs at
$\nu_{peak}$, the peak of the black body distribution.}

\item{We assume that in the jet frame equipartition holds between the
energy densities of the relativistic particles and the magnetic field.
This allows us to estimate the average field strength in the source.
Moreover, we take the volume filling factor to be one and the
contribution to the particle energy density by protons to be nil.  If
relativistic protons contribute significantly to the particle energy
density, B$_{eq}$ increases and beaming parameters become more extreme
(larger values of $\Gamma$ and smaller angles).}

\item{The spectral index measured for the synchrotron spectrum at the
lowest available frequencies continues unchanged to frequencies much
lower that those accessible to Earth based telescopes.  The gross
attributes of this extrapolation must be consistent with radio
emission arising from $\gamma'_{\rm e} \approx 2000$ and X-ray
emission (0.2-10 keV) arising from $25 < \gamma'_{\rm e} < 300$.}
This extrapolation of the electron spectrum provides a large
uncertainty in the beaming model and has been tacitly ignored by most
previous formulations.  At our current level of understanding, we do
not have any convincing evidence on the amplitude or the power law
index of the electron distribution at low energies.
 
\end{itemize}

\subsection{Beaming parameters for a number of sources}

In this section, we provide some of the key synchrotron parameters and
beaming descriptors for a few knots in radio jets.  For the synchrotron
parameters, we use the standard expressions (e.g. Pacholczyk, 1970)
with observables transformed back to the jet frame.  For the beaming
parameters, our solution to eq.~\ref{eq:result} requires only 4
numbers: 

\begin{itemize}

\item{B(1), the equipartition field from the synchrotron model
when $\delta$=1 (no beaming)}

\item{R(1), the ratio of amplitudes of the power law spectra:
$k_{ic}/k_{sync}$ where both amplitudes must refer to the same spectral
index}

\item{z, the redshift}

\item{$\alpha$, the spectral index}

\end{itemize}

\begin{figure}
\plotone{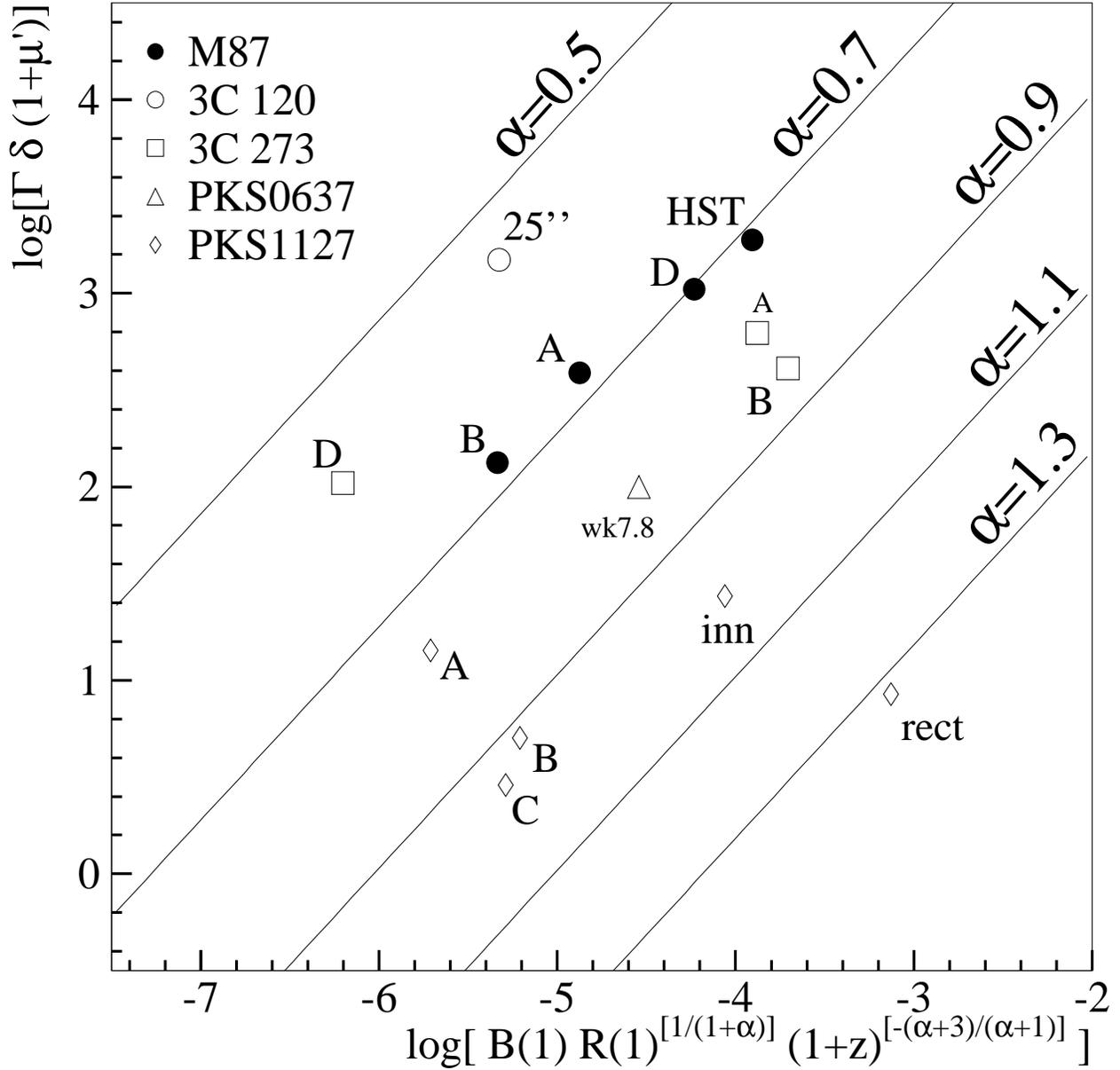}
\caption{A graphic illustration of eq.~\ref{eq:resulta}.
Since the second $\Gamma$ term of eq.~\ref{eq:result} has been
dropped, the results shown here are indicative only. Solutions for
several values of the spectral index are shown by the solid lines.
The source data from Table~~\ref{tab:results} demonstrate that
steeper spectra and higher redshifts require less beaming than do
lower redshift, flatter spectra knots.  Of the four
'observables' contributing to the X axis value, the equipartition
field for no beaming, B(1) varies by a factor of 10, whereas R(1)
varies by at least 9 orders of magnitude (mitigated here, of course,
by the exponent which is usually less than unity).\label{fig:results}}
\end{figure}

Table \ref{tab:results} gives the results and it can be seen that
the beaming parameters range from quite modest values (e.g. PKS1127)
to rather unbelievable extremes (3C 120).  We have plotted the key
results in fig.~\ref{fig:results} which is a representation of the
beaming parameters as a function of the observables (eq.~\ref{eq:resulta}).



\section{Evaluation of beaming and synchrotron models}\label{sec:conflict}

In this section, we deal with the conflicting evidence for general
beaming models and for synchrotron models.  On the one hand, some sort
of beaming appears to be required by the observation that all of the
known jet sources (excluding of course the SSC terminal hotspots)
produce X-ray emission on only one side, and that is the side which
has the only or dominant radio jet and for which relativistic effects
have been demonstrated (usually on VLBI scales).  On the other hand,
for knots such as A1/3C273 or B/3C390.3, the observed X-ray intensity
is accommodated by an extension of the power law connecting the radio
and optical (i.e. the synchrotron spectrum).  If the X-ray emission
from these knots were to be IC/CMB emission enhanced by beaming, this
situation would be only a coincidence.

\subsection{X-ray spectra}\label{sec:xspec}

In general, we expect synchrotron spectra to be concave downward when
plotted on the usual log S vs. log $\nu$ plane.  For high energy
electrons, this is caused by the E$^2$ losses.  Although we do not
have reliable observational evidence on the shape or amplitude of
the electron spectrum for $\gamma~\leq~500$, our expectation is that
the spectrum observed in the radio regime might flatten at lower
frequencies, but it is unlikely to steepen.  This view is dictated by the
requirement to avoid divergences in the total number of electrons and
integrated energy in the electron spectrum.

Reliable X-ray spectra are now available for only a few jets, but so
far, the evidence is that the spectra are either the same as observed
in the radio/optical (e.g. 3C 273/A1, Marshall et al. 2001a), or
steeper ($\alpha_x$(M87)=1.5$\pm$0.4, B\"{o}hringer et al. 2001 and
$\alpha_x$(3C66B)$\approx1.3$, Hardcastle et al. 2001b).  This
supports the synchrotron model and is not in accord with our
expectation for beaming from low energy electrons.  However, a
relatively hard spectrum is found for PKS1127 ($\alpha_x$=0.5$\pm$0.2;
Siemiginowska et al. 2001), a jet that requires very modest
beaming parameters.  Similarly, Chartas et al. (2001) find
$\alpha_x$=0.85$\pm$0.1 for the outer jet in PKS0637, another likely
beamed IC source.  The X-ray spectral index may prove to be a key
parameter in differentiating synchrotron from beaming sources.
$\alpha_x\leq\alpha_r$ may be indicative of beaming whereas
$\alpha_x>\alpha_r$ is the condition we expect from homogeneous
synchrotron models.

\subsection{Distinct spectral components}\label{sec:distinct}

By this we mean a population of relativistic electrons which is
distinct from that we measure in the radio/optical regime.  The
emitting volume could be quasi coincident with that of the known
synchrotron component, or it could be embedded within the known
volume.  For synchrotron models, a small, flat spectrum component is
required. In the case of the 25'' knot in 3C 120, the current optical
upper limits demand that $\alpha~\leq~0.4$.  A small region within the
radio knot would be an efficient method to produce the observed X-rays
in the sense that the total energy of such an entity would be very
modest (Harris et al. 1998).  As noted in sec~\ref{sec:xspec}, the
spectral data so far available do not support this sort of scenario.

For the IC/CMB model without beaming, what is required is an excess
population of electrons with $\gamma$ in the range 500-2000.  For bulk
relativistic velocities, this energy range shifts to lower values
since the peak of the CMB appears blue shifted to the jet.

\subsection{Jet length and structure}

The beaming model has untenable consequences for straight jets.  In
general, the angle, $\theta$, between the jet axis and line of sight
to the observer derived for jets such as 3C 273 are not very different
from those found at VLBI scales, and this jet appears straight on the
kpc scale.  If the jet fluid moves in a straight line along the axis
of the jet, then the physical length of the jet is the projected
length divided by sin~$\theta$.  In Table~\ref{tab:results} we have
listed these values which range up to a Mpc for 3C 273.  Very few
radio sources are directly observed (i.e. in the plane of the sky) to
have such large physical sizes.

These excessive lengths can be avoided in at least two ways.  We know
that some jets bend (e.g. 3C 120).  If the apparently straight jets
(e.g. 3C 273) were to bend significantly in a plane normal to
the plane of the sky, they would appear straight but the beaming
required for a particular feature would not apply to the whole jet.
The other possibility is that the jet fluid responsible for the
observed radiation would not be constrained
to move along the jet axis, but, for example, might follow a helical
path (see e.g. Meier, Koida, and Uchida, 2001) around the axis as
suggested by the HST images of 3C 273, or the emission regions might
represent all or a portion of the flow being deflected toward the
observer.

Whichever of these possibilities might be operative, it should work in
conjunction with the requirement that the jet seen is the one
approaching the observer.

\subsection{Evidence for bulk relativistic velocities and cold jets}

Are there reasonable expectations for kpc scale relativistic
velocities aside from their utility in explaining X-ray emission?
Bridle (1996) marshals considerable evidence for bulk relativistic
velocities at least for the inner section of FRI jets, and all the way
out to the terminal hotspots for FRII radio galaxies and quasars.
Although representative values for such velocities are not presented,
we infer $\Gamma$ values of at least a few.  Wardle \& Aaron (1997),
while agreeing that one-sidedness evidence supports relativistic
velocities in jets, argue that the observed dispersion in the ratio of
the flux density of the inner straight part of the jet (the feature
expected to be most enhanced by beaming) to the flux density of the
lobe (no beaming) is less than the expected dispersion for
$\Gamma~\approx~5$.  Thus they conclude that the characteristic jet
speed is $\beta~\approx$~0.85 ($\Gamma~\approx~2$) and that all jets
decelerate between pc and kpc scales.

The consideration of large $\Gamma$ values for kpc scale jets reminds
us of the problem of energy transport.  If electrons were to be a
significant component of the jet, they cannot escape the IC losses
which increase as $\Gamma^2$.  Since these losses go as $\gamma^2$, we
may expect that only low energy electrons survive for long distances.
Thus the energy must be transported by Poynting flux, protons, or cold
electrons and positrons.  If the latter, it might be possible to
explain the relatively smooth X-ray emission such as that underlying
the 3C 273 knots or the Pictor A inner jet, as IC beamed emission from
essentially cold electrons which are characterized by
$\Gamma~\geq~\gamma$.  The brighter knots would then rely on the
conventional shock acceleration and/or pair production to generate 
(in situ) the usual power law distribution of relativistic electrons
with $\gamma$ values of normal synchrotron sources.


\subsection{Offsets between radio and X-ray brightness peaks}

The synchrotron expectation is that lower frequency emission will persist
further downstream since the higher energy electrons have shorter
lifetimes.  However, at the site of the shock, there should be strong
emission at all frequencies (e.g. M87, knot A).  In the beaming model,
one way to accommodate a leading X-ray feature is to argue for the
existence of a precursor shock which is incapable of producing
electrons with $\gamma~\geq~1000$.

\subsection{Progression of brightnesses for multiple knot jets}

In the optical, the 3C 273 jet has a series of well defined knots with
comparable brightnesses and fluxes.  However, at radio wavelengths,
the knot intensities increase going out the jet and the situation is
reversed for X-rays.  This X-ray and radio behavior is mimicked in
PKS1127.  As a natural consequence, the parameters for the beaming
model become less restrictive (smaller $\Gamma$, larger $\theta$) as
the distance from the core increases.

This sort of behavior is consistent with synchrotron models for which
we might well expect that the properties of shocks which permit them to
produce copious supplies of the highest energy electrons will diminish
as we move out to larger distances.

\subsection{Is overproduction a problem?}

The beaming model means that the underlying synchrotron components are
significantly less energetic than for the stationery synchrotron
model.  Typical luminosities drop from 10$^{41}$ erg~s$^{-1}$ to values like
10$^{38}$~erg~s$^{-1}$.  Such entities must be very common in jets, so might
we expect to be overwhelmed with X-ray emission from excessive numbers
of weak synchrotron emitters which happen to be moving towards us?

\section{Conclusions}

There is little doubt in our minds that thermal bremsstrahlung is not
a major contributor to the X-ray emission from most of the jet
features discussed in this paper.  We find the arguments in favor of
SSC emission from terminal hotspots of a few FRII radio galaxies to be
convincing.

The failure of simple synchrotron models to fit the spectra of some
features, together with the quasi-universality of one-sidedness
almost demands the occurrence of relativistic bulk velocities.  On the
other hand, extreme values for the beaming parameters together with
the excellent agreement of the X-ray spectra with the extrapolation of
the radio-optical spectra for many features lend strong support to the
synchrotron models.  

Thus the leading contender for most of the X-ray jets appears to be a
single or multiple spectral component  synchrotron model augmented by
a modest beaming with $\Gamma~<~10$ to account for the one-sidedness
of most of the recorded X-ray features.  The IC/beaming hypothesis can
consistently explain the emission from several jets, and the low
required beaming factors for some sources as e.g.\ PKS~1127 make it
probable that this emission component contributes substantially to the
observed X-ray fluxes of at least some of the observed X-ray features.
Higher redshifts and steeper radio spectra appear to favor beaming,
while local sources with flatter radio spectra are probably dominated by
synchrotron emission.




\acknowledgments

DEH thanks W. Tucker for extensive discussions during the early phases
of this investigation, T. Kneiske for discussions on comparison of
beaming formulations, and A. Bridle for discussing the radio aspects
of jets.  We also thank M. Hardcastle, M. Birkinshaw, and R. Sambruna
for communicating results prior to publication and A. Siemiginowska
and H. Marshall for sharing their proprietary data.  M. Birkinshaw,
D. Worrall, W. Forman, F. Aharonian, and A. Marscher provided helpful
suggestions for improving the manuscript and the anonymous referee
supplied useful criticisms.  HK thanks P. Coppi for discussions on AGN
jets, and acknowledges support by NASA (NAS8-39073 and GO 0-1169X).
The work at SAO was partially supported by NASA contract NAS8-39073.
 


\appendix

\section{Formulation of bulk relativistic beaming with equipartition fields}

We first find the synchrotron parameters in the jet frame and then
derive expressions for the IC emission.  Then we obtain two
expressions for the ratio of IC to synchrotron losses: one based on
the observed data and one based on the energy densities of phtons and
magnetic fields.  Equating these two expressions provides the final
beaming equation.  We use cgs units throughout except where stated
otherwise.
 
The basic relationships for the parameters which describe the bulk
motion of the jet fluid are the jet velocity,
$\beta_{\rm j}~\equiv~\frac{v}{c}$; the Lorentz factor, $\Gamma$; the
angle of the fluid's velocity with respect to the line of sight of the
observer, $\theta_{\rm j}$; and the Dopler beaming factor, $\delta$.

\begin{equation}
\Gamma~\equiv~\frac{1}{\sqrt{1~-~\beta^2}}
\label{eq:gamma1}
\end{equation}

\begin{equation}
\delta^{-1}~=~\Gamma(1~-~\beta~cos~\theta)
\label{eq:delta1}
\end{equation}

\subsection{The synchrotron parameters in the jet frame} 
The required observables are:

\begin{itemize}

\item{the size of the emitting volume}

\item{$\alpha$, the low frequency spectral index of the synchrotron spectrum.}

\item{$\nu_1$ and $\nu_2$, the lower and upper limits of the
synchrotron spectrum in the observer frame.  NB: It may be necessary
to decrease $\nu_1$ to $\approx$~1~MHz so as to include low $\gamma$
electrons required for the beaming model.}

\item{$S_{\rm o}$ at some $\nu_{\rm o}$, with $S_{\rm o}$ the flux
density observed at the Earth, and $\nu_{\rm o}$ the frequency at the
receiver corresponding to $S_{\rm o}$.  This provides $k_s$, the
amplitude of the observed synchrotron spectrum: $k_s = S_{\rm
o}~\nu_{\rm o}^{\alpha}$.}
 
\item{The luminosity distance and redshift.}

\end{itemize}

We then move to the jet frame.  Frequencies convert as:

\begin{equation}
\nu'=\nu\, (1+z) / \delta
\label{eq:shiftf}
\end{equation}

Assuming a knotty jet and neglecting complications arising from
different pattern and jet plasma velocities, we find from the Lorentz
invariant, $S/\nu^3$ (see also eqs. C7 and C11 of Begelman, Blandford,
\& Rees 1984), the monochromatic luminosity per solid angle in the jet
frame:

\begin{equation}
l'_{\Omega,\nu'} = \frac{k_s(obs)~\nu'^{-\alpha} 
(1+z)^{\alpha - 1}~D_{\rm L}^2}{\delta^{3+\alpha}}~~{\rm~erg~cm^{-2}~s^{-1}~Hz^{-1}~str^{-1}}
\label{eq:shifts}
\end{equation}

\noindent 
Here, $D_{\rm L}$ is the luminosity distance and $(1+z)^{\alpha-1}$ is
the $K$-correction.  This equation allows evaluation of
$l'_{\Omega,\nu'}$ at any convenient frequency in the jet frame.
Integrating over frequency and solid angle, the total luminosity
becomes:

\begin{equation}
L_s'\, = \,\left[\frac{4\pi D_{\rm L}^2 }{\delta^4}\right]
\left[\frac{k_s\,(\nu_2^{1-\alpha} - \nu_1^{1-\alpha})}
{(1-\alpha)}\right]\, = \,\frac{L_s}{\delta^4}~{\rm~erg~s^{-1}}
\label{eq:lum}
\end{equation}

The right term is the energy flux at the Earth, $4\pi~D_{\rm L}^2$
computes the luminosity at the source, and $\delta^4$ transforms into
the jet frame.

To compute the equipartition magnetic 
field in the jet frame (e.g. eq. 7.12 of Pacholczyk, 1970):

\begin{equation}
B'_{\rm eq} = \left[\frac{18.85\, C_{12}\, (1+k)\, L_s'}{\phi V'}\right]^{2/7}~~{\rm~Gauss}
\label{eq:beq}
\end{equation}

\noindent
V' is the emitting volume; $\phi$ is the volume filling factor; and
$C_{12}$ is a Pacholczyk parameter which is a slowly varying function
of $\alpha, \nu'_1$, and $\nu'_2$ (e.g. $C_{12}$=5.7$\times 10^6$ for
$\alpha=0.8; \nu'_1=10^7; \nu'_2=10^{15}$).  When the correct
expression for $C_{12}$ is used in eq.~\ref{eq:beq}, it introduces an
extra factor of $\delta^{0.5}$ to the numerator within square
brackets, so that together with the explicit $\delta^4$ from
eq.~\ref{eq:lum} which appears in the denominator, the final
dependency within the square brackets goes as $\delta^{-\frac{7}{2}}$
and hence:

\begin{equation}
B'_{\rm eq} = \frac{B(1)}{\delta}
\label{eq:b1}
\end{equation}
 
\noindent
where B(1) is the equipartition field calculated for no beaming
($\delta$=1).

\subsection{Converting angles to the jet frame in order to calculate
the IC emission}

From manipulation of the basic equations (\ref{eq:gamma1},~\ref{eq:delta1}):

\begin{equation} 
\mu_{\rm j} \,\equiv\, \cos{( \theta_{\rm j})}\, = \, 
\frac{ \Gamma - 1/\delta}{\sqrt{\Gamma^2 - 1}}
\label{eq:mu}
\end{equation}

Since in the jet frame most of the CMB photons will come from
the direction in which the jet plasma is moving and the scattered IC
emission from a particular electron will be strongly beamed into the
direction of the instantaneous velocity vector of the electron, the IC
emission will not be isotropic in the jet frame.  Take the angle
between the l.o.s. to the observer and the jet axis as $\theta_{\rm
j}$ and $\theta'_{\rm j}$ in the observer and jet frames,
respectively. Likewise, $\mu_{\rm j}$ and $\mu'_{\rm j}$ are the
corresponding cosines. Then:

\begin{equation}
\mu_{\rm j} '\, = \,\frac{\mu_{\rm j} - \beta_{\rm j}}{1-\mu_{\rm j}\, \beta_{\rm j}}
\label{eq:shiftmu}
\end{equation}

where $\beta_{\rm j}$ is the velocity of the jet (see e.g. Pacholczyk 1970,
eq. 5.25).  The inverse is also useful:

\begin{equation}
\mu_{\rm j}\, =\, \frac{\mu'_{\rm j} + \beta_{\rm j}}{1 +\mu'_{\rm j}\, 
\beta_{\rm j}}
\label{eq:shiftmup}
\end{equation}

To calculate the power emitted by an electron traveling at an angle
$\Theta'$ to the seed photon direction, we use $\Theta'\,\approx$
$\theta'_{\rm j}$ and follow Rybicki \& Lightman (1979, p. 199, see
also Dermer et al. 1992) to get:

\begin{equation}
\frac{dE'}{dt} = c \,\sigma_T\, u_{\rm CMB}'\, \kappa~~{\rm~erg~s^{-1}}
\label{eq:power}
\end{equation}

\begin{equation}
\kappa = \gamma_{\rm e}'^2 (1 + \beta'_{\rm e}\, \mu'_{\rm j})^2 - 
(1 + \beta'_{\rm e}\, \mu'_{\rm j})
\label{eq:kappa}
\end{equation}

\noindent where $\sigma_{\rm T}$ is the Thomson cross section; $E'$ is
the energy of the electron; $u'_{\rm CMB}$ is the Doppler boosted
energy density of the CMB; and $\gamma'_{\rm e}$, $\beta'_{\rm e}$ are
the Lorentz factor and velocity of the electron in the jet frame.

The second term in eq.\ref{eq:kappa} will be negligible except when
$\mu'_{\rm j}$ approaches minus one, which is the case for electrons
moving directly away from the jet's velocity vector.

The energy density of the CMB (as seen by the jet moving with $\Gamma$):
 
\begin{equation}
u'_{\rm CMB}\, = 4 \times 10^{-13}\, (1+z)^4\, (4/3)\,
(\Gamma^2 - 1/4)~~{\rm~erg~cm^{-3}}
\label{eq:unu}
\end{equation}
 
The peak frequency of the CMB (again, as seen by the jet)
 
\begin{equation}
\nu'_{\rm p} = 1.6 \times 10^{11}\, (1+z)\, \Gamma~~{\rm~Hz}
\label{eq:nup}
\end{equation}

In the jet frame, the mean energy of photons scattered in the
direction $\theta'_{\rm j}$ will be:

\begin{equation}
<\varepsilon'_{\rm IC}> = \frac{dE'/dt'} {\left[c\, \sigma_{\rm T}\, n'_{\rm CMB}\,
(1 + \beta'_{\rm e}\, \mu'_{\rm j})\right]}~~{\rm~erg}
\label{eq:icpower1}
\end{equation}

\begin{equation}
\mbox{\hspace*{1cm}} =\, \frac{<\varepsilon'_{\rm p}> \kappa}{1 + \beta'_{\rm
e}\, \mu'_{\rm j}}~~{\rm~erg}
\label{eq:icpower2}
\end{equation}

\begin{equation}
\mbox{\hspace*{1cm}}=\,<\varepsilon'_{\rm p}> \left[\gamma_{\rm e}'^2 (1 + \beta'_{\rm e}\, 
\mu'_{\rm j}) - 1\right]~~{\rm~erg}
\label{eq:icpower3}
\end{equation}

\noindent 
where $<\varepsilon'_{\rm p}>$ is the mean energy of the photons before scattering: $1.6
\times 10^{11} (1+z) \Gamma$, and $n'_{\rm CMB}$ is the seed photon
density in the jet frame, so that the denominator in
Eq. (\ref{eq:icpower1}) is the (expected) number of scatterings per
unit time for one electron.

\subsection{The expression for the observed ratio of luminosities, R(obs)}
 
We define R as the ratio of IC and synchrotron luminosities.  Since
both of these losses goes as the electron energy squared, all
electrons will experience the same value of R.


For $\alpha$ = 1,
there is equal energy per decade and thus the integration band is not
important so that the ratio of luminosities is equal to the ratio of
amplitudes of the X-ray (IC) and radio (sync) spectra.

\begin{equation} 
R(1) = \frac{L_{ic}}{L_{sync}} = \frac{k_x}{k_r}
\label{eq:r1}
\end{equation}

For $\alpha \neq 1.0$, to obtain $L_{ic}/L_{sync}$ we change
integration parameters from $\nu'$ to $\gamma'$.  These conversions
are found by using Eq.\ (\ref{eq:icpower3}) with $\beta'_{\rm e}$=1 and 
dropping the '-1':

\begin{equation}
\nu'_{\rm ic} = (1+\mu'_{\rm j})\, \gamma_{\rm e}'^2\, [1.6\,\times\, 
10^{11}\, (1+z)\,\Gamma]~~{\rm~Hz}
\label{eq:fic}
\end{equation}
\noindent
(the last term is the peak frequency of the CMB in the jet frame)

\begin{equation}
\nu'_{sync} = 4.202\,\times\, 10^{6} \,\gamma_{\rm e}'^2\, \times\, B'~~{\rm~Hz}
\label{eq:fsy}
\end{equation}

Then most of the factors of eq.~\ref{eq:lum} including the integration
itself, cancel and the remaining parameters are those resulting from
the different conversions of $\nu$ to $\gamma$.  After these
manipulations, we find:

\begin{equation}
R(obs) = R(1)\left[\frac{3.808~\times~10^4\, (1+z)\, (1+\mu'_{\rm
j})\, \Gamma}{B_{\rm eq}'}\right]^{1-\alpha}
\label{eq:robs}
\end{equation}

\subsection{Evaluate R from ratio of energy densities}

The ratio of the IC and synchrotron luminosities can be computed from
the ratio of energy densities in photons and magnetic field.  This
provides an expression for the 'expected' value of R.  In using
eq.(\ref{eq:kappa}), we have dropped the second term since it will
normally be much smaller than the first term.

\begin{equation} 
R(exp)\, = \,\frac{U'_{\rm CMB}}{u'(B_{\rm eq}')}
, =\, \frac{4\, \times\, 10^{-13}\, (1+z)^4\,
(1+\mu'_{\rm j})^2\, \left[\Gamma^2-(1/4)\right]}{B_{\rm eq}'^2/8\pi}
\label{eq:rexp}
\end{equation}

\noindent
where $U'_{\rm CMB}$ is the CMB energy density in the jet frame, 
modified by the deviation from the beamed emission power
from that computed for an isotropic seed photon field.

\subsection{The beaming equation}
 
Equating the two expressions for R (Eqs. \ref{eq:robs} and \ref{eq:rexp}),

\begin{equation}
\Gamma^{\alpha+1} - \frac{\Gamma^{\alpha-1}}{4}\, =\,
R(1)\,\frac{9.947\, \times\, 10^{10}\,
(3.808~\times~10^4)^{1-\alpha}}{(1+z)^{3+\alpha}}\left[\frac{B_{\rm
eq}'}{1+\mu'_{\rm j}}\right]^{\alpha +1}
\label{eq:result}
\end{equation}

Thus with 4 parameters which depend on the observable data ($\alpha$,
z, B(1), and R(1)) we can solve eq.~\ref{eq:result} numerically by
finding the range  of $\mu'$ and $\Gamma$ pairs which satisfy the equation.

If we ignore the second $\Gamma$ term and replace B$^{\prime}$ with
eq.~\ref{eq:b1}, we obtain an approximate expression for the beaming
parameters in terms of the observables (see~fig.~\ref{fig:results}):


\begin{equation}
\Gamma\delta(1+\mu'_{\rm j})=\frac{B(1)[10^{11}R(1)]^{\frac{1}{1+\alpha}}
(3.808~\times~10^4)^{\frac{1-\alpha}{1+\alpha}}}{(1+z)^{\frac{\alpha+3}{\alpha+1}}}
\label{eq:resulta}
\end{equation}

\section{Frequencies, electron energies, and halflives}

Once we have the allowed values of $\theta_{\rm j}, \delta,$ and
$\Gamma$, we need to determine where the emission bands occur and
which segments of the electron energy spectrum are responsible for the
observed radiation.

To determine the IC emission frequency of electrons responsible for a
particular synchrotron frequency:

\begin{equation}
\nu_{\rm ic}(obs) = \nu_s(obs)\, 3.808~\times~10^4\, (1+\mu'_{\rm
j})\, \Gamma (1+z) / B'~~{\rm~Hz}
\label{eq:nuic}
\end{equation}

and the reverse:

\begin{equation}
\nu_s(obs) \,= \,\frac{\nu_{\rm ic}(obs) \,2.626\,\times 10^{-5}\,
B'}{\Gamma (1+z) (1+\mu'_{\rm j})}~~{\rm~Hz}
\label{eq:nus}
\end{equation}

To find the electron energy responsible for a particular synchrotron
frequency, we use eqs~\ref{eq:shiftf},~\ref{eq:b1}, and \ref{eq:fsy}:

\begin{equation}
\gamma_{\rm e}'^2\, =\, \frac{2.380\times 10^{-7}\, \nu_s(obs)\, (1+z)}{B'\, \delta}
\label{eq:gammas}
\end{equation}

For the electron energy responsible for an IC frequency we use
eqs~\ref{eq:shiftf} and \ref{eq:fic}:

\begin{equation}
\gamma_{\rm e}'^2 \,=\, \frac{6.25\times 10^{-12}\, \nu_{\rm ic}(obs)}{(1+\mu'_{\rm j}) \delta \Gamma}
\label{eq:gammaic}
\end{equation}

For the halflives, we use the normal expression: $(dE/dt) \times \tau = E/2$.
This results in:

\begin{equation}
\tau'\,=\,\frac{10^{13}}
{\gamma'_{\rm e} 
 \left\{
  1.016\,B'^2\, + \,10.398\,(1+z)^4 (\Gamma^2-\frac{1}{4}) 
  \left[(1+\beta'_{\rm e} \mu'_{\rm j})^2 - 
     \frac{(1+ \beta'_{\rm e} \mu'_{\rm j})}
          {\gamma_{\rm e}'^2}
  \right] 
 \right\} }
\label{eq:tauj}
\end{equation}

\noindent
where $B'$ is in $\mu$G and $\tau'$ is in years.

Making the usual approximations for quantities close to one, and since
time intervals in the jet frame are observed at the Earth 
as:

\begin{equation}
\tau_{\rm o} = \tau' \,\frac{(1+z)}{\delta}
\label{eq:tauo1}
\end{equation}

an approximate expression for the observed halflife is:
 
\begin{equation}
\tau_{\rm o}\, =\, \frac{10^{13}\,(1+z)}{\gamma'_{\rm e}\, \delta\, \left\{B'^2 \,+\, 40\,(1+z)^4
\Gamma^2\right\}}~~{\rm~yrs}
\label{eq:tauo2}
\end{equation}



\section{References}


Aharonian, F.A. 2001, MNRAS (in press) astro-ph/0106037

Begelman, M., Blandford, R., \& Rees, M. 1984, Rev. Mod. Phys. 56, on page 340, Eqs. C7 and C11

Biretta, J.A., Stern, C.P., and Harris, D.E. 1991 A.J., 101, 1632 (M87)

Bodo, G., Ghisellini, G., \& Trussoni, E. 1992, MNRAS, 255, 694-700

B\"{o}hringer, H. et al. 2001, A\&A 365, L181-L187 (M87)

Bridle, A.H. 1996 ASP Conference Series 100, 383-394, ``Energy
Transport in Radio Galaxies'', Hardee, Bridle, \& Zensus, eds.

Brunetti, G., Cappi, M., Setti, G., Feretti, L., Harris, D.E. 2001,
A\&A 372, 755

Burbidge, G.R. 1956, ApJ 124, 416-429

Carilli, C.L., Perley, R.A., \& Harris, D.E. 1994, MNRAS 270, 173-177
(Cyg A)

Celotti, A., Ghisellini, G., \& Chiaberge, M. 2001 MNRAS 321, L1-5

Chartas, G. et al. 2000 ApJ 542, 655 (PKS0637)

Dermer, C.D, Schlickeiser R., Mastichiadis A.\ 1992, A\&A, 256, L27

Ginzburg, V.L. and Syrovatskii, S.I. 1969 ARA\&A 7, 375

Hardcastle, M.J., Birkinshaw, M. \& Worrall, D.M. 2001a MNRAS
(accepted; 3C 123 astro-ph/0101240 )

Hardcastle, M.J., Birkinshaw, M. \& Worrall, D.M. 2001b MNRAS
(submitted;  3C 66B astro-ph/0106029)  

Harris, D.E., Carilli, C.L. and Perley, R.A. 1994 Nature 367, 713 (CygA)

Harris, D.E., Leighly, K.M., and Leahy, J.P. 1998 ApJ 499, L149-L152 (3C390.3)

Harris, D.E., Hjorth, J., Sadun, A.C., Silverman, J.D. \& Vestergaard,
M. 1999 ApJ 518, 213-218 (3C120)

Harris et al. 2000 ApJ 530, L81 (3C295)



Kraft, R.P. et al. 2000 ApJ 531, L9-L12 (Cen A)

Mannheim, K., Krulls, W.M., and Biermann, P.L. 1991 Astron Astrophys
251, 723

Marshall, H. et al. 2001a ApJ 549, L167 (3C 273)

Marshall, H. et al. 2001b ApJ (submitted) (M87)

Meier, D.L., Koide, S., \& Uchida, Y. 2001, Science 291, 84-92
  
Pacholczyk, A.G. 1970, ``Radio Astrophysics'' W. H. Freeman , San
Francisco
 
Readhead, A.C.S. 1994 ApJ 426, 51-59

Rybicki, G.B. \& Lightman, A.P. 1979, ``Radiative Processes in
Astrophysics'' John Wiley \& Sons, New York

Sambruna, R.M., Urry, C.M. Tavecchio, F., Maraschi, L., Scarpa, R.,
Chartas, G., \& Muxlow, T. 2001 ApJ 549, L161 (3C 273)

Schreier, E.J., Gorenstein, P., \& Feigelson, E.D. 1982, ApJ 261, 42

Siemiginowska, A.L. et al. 2001 (in preparation) (PKS1127-145)

Singal, A. K. 1986, A\&A, 155, 242-246

Tavecchio, F., Maraschi, L., Sambruna, R.M., \& Urry, C.M. 2000, ApJ
544, L23

Wardle, J.F.C., \& Aaron, S.E. 1997, MNRAS 286, 425-435

Wilson, A.S., Young, A.J., and Shopbell, P.L. 2001a, ApJ 544, L27 (CygA)

Wilson, A.S., Young, A.J., \& Shopbell, P.L. 2001b, ApJ 547, 740 (PicA)


\clearpage





\clearpage


\begin{table}
\scriptsize

\caption{\textbf{2001 March list of radio sources with jet related X-ray
emission}}\label{tab:sources}
\bigskip

\begin{tabular}{lcclrccccl}

Generic	&RA	&Dec	 & z	 &Dist.	&kpc/"   &Assoc. 	&Assoc. 	&PA  	&Classification \\
Name	&J2000	&J2000	&	&(H=50)  &(H=50)  &radio	&optical	&w.r.t. &\\      
	&hh mm	&dd mm	&	 &(Mpc)  &        &     	&		&core	&        \\
3C 15 	&00 37  &-01 09 &0.0730 &453    &1.91   &knots  &knots  &  -30&nyc\\
3C 31   &01 07  &32 25  &0.0167 &101    &0.47   &jet    &no     &  -20   &nyc\\
3C  66B &02 23  &43 00  &0.0215 &130    &0.61   &jet    &jet    &   45    &sync\\
3C 120	&04 33	 &05 21	&0.0330	  &201	 &0.91 	&knot	 &no	&NW	&beam?\\
3C 123	&04 37	 &29 40	&0.2177	 &1448	 &4.74	 &hs	 &no	 &110	&SSC \\
PictorA	&05 19	&\llap{-}45 46	&0.0350	 & 214	 &0.97	&W hs	 &yes	 &-80	&beam\\
PKS0521 &05 25  & -36 27 	&0.055   &338    &1.475 &knots   &yes    &NW      &nyc\\
PKS0637	&06 35	&\llap{-}75 16	&0.653	& 5197	 &9.22	&knots	 &yes	 &-90	&beam \\
PKS1127 &11 30	&\llap{-}14 49	&1.18	&11257	&11.48	&yes	  &?  	  &42	&dis.morph.\\
3C 263  &11 39  &65 47  	&0.6563  &5230  & 9.25  &yes    &no?     &116   &SSC\\
3C 273	&12 29	 &02 02		&0.1583	 &1025	 &3.70	&knots	&knots	 &190	&sync,beam \\
NGC4261	&12 19	 &05 49		&0.00737 &44    &0.21   &yes	& no?	 &-90	&nyc \\
M87	&12 30	 &12 23		&0.00427 & 16	 &0.077  &knots	&knots	 &-60	&sync \\
Cen A	&13 26	&\llap{-}42 49	&  	&  3.5  &0.017	 &?	 & ?	 & 70	&dis.morph.\\
3C 295	&14 11	 &52 13		&0.45	& 3307	 &7.63	 &2hs	 &yes	& -10	&SSC \\
3C 371	&18 06	 &69 49		&0.051	 &314	 &1.38	 &yes	 &yes	&WSW	&sync \\
3C390.3 &18 42	 &79 46		&0.0561	  &346	 &1.50	 &hsB	 &hsB	&-10	&sync \\
Cyg A	&19 59	 &40 44		&0.0560	  &345	 &1.50	 &2hs	 &no	&110/280 &SSC \\
\end{tabular}


\bigskip
Notes and Comments

$q_{\circ}$ = 0

References for individual sources as well as contact persons for
unpublished data can be found at the XJET website:
http://hea-www.harvard.edu/XJET/

In the classification column, `nyc' means ``not yet classified'' (the
data are unpublished) and `dis.morph.' signifies ``discrepant
morphology''.

Morphology: we generally use 'knot' to indicate a distinct brighter feature in 
a jet that continues past the feature and 'hotspot' either as the terminal
bright enhancement at the end of an FRII jet, or as one of the multiple
features associated with the termination of a jet.  Normally 'knots' are 
found on the inner portion of FRI jets whereas 'hotspots' are mostly
at the ends of FRII jets.  However, we are not trying to impose distinct
definitions, and infer no physical differences beyond these generalities.

\end{table}
\clearpage

\small

\begin{table}
\scriptsize
\caption{\textbf{Beaming parameters for selected jet
features}}\label{tab:results}
\bigskip

\begin{tabular}{llllrl|ccr|ccr} \hline \hline
&&\multicolumn{4}{l}{INPUT PARAMETERS}&\multicolumn{3}{|c|}{BEAMING}&\multicolumn{3}{c}{JET LENGTH} \\
&&&&&&\multicolumn{3}{c|}{PARAMETERS}&&& \\
Source&Knot&z&$\alpha$&B(1)&R(1)&$\Gamma$&$\theta$&R$^{\prime}$&DfC&proj&phys \\
&&&&($\mu$G)&&$\delta$&(deg)&&(arcsec)&(kpc)&(kpc) \\ \hline
&&&&&&&&&&& \\
M87&HST-1&0.00267&0.71&272&0.264&40&1.2&551&1&0.08&4 \\
       & D&...&0.70&319&0.0568&33&1.8&123&3&0.23&7 \\
       & A&...&0.67&288&0.006&19&2.8&21&12&0.92&19 \\
       & B&...&0.67&248&0.0013&11&4.7&3&15&1.15&14 \\
&&&&&&&&&&& \\  							 
3C 120 & 25"&0.033&0.54&41&0.04&39&1.5&15764&25&23&869 \\
&&&&&&&&&&& \\		 
3C 273 & A1&0.158&0.79&172&1.09&25&2.3&246&13&48&1196 \\
       & B1&...&0.85&134&3.5573&20&2.8&166&17&62&1269 \\
       & D/H&...&0.57&221&0.00017&10&5.5&5&20&75&783 \\
&&&&&&&&&&& \\									 
PKS0637& wk7.8&0.651&0.81&195&0.209&9.9&5.7&21&8&72&725 \\
&&&&&&&&&&& \\										 
PKS1127& inner&1.187&1.0&144&8.426&5.2&11&8&1.5&17&89 \\
       & rect.&...&1.33&24&8.9E4&2.9&20&31&11&126&368 \\
       & A &...&0.76&13&0.667&3.7&15&287&12&138&533 \\
       & B &...&0.92&29&1.106&2.2&26&7&19&221&504 \\
       & C &...&0.94&29&0.754&1.5&32&3&28&321&606 \\
&&&&&&&&&&& \\ \hline
\end{tabular}
\end{table}

\clearpage
\small
\begin{table}
\scriptsize
\begin{tabular}{llccc|cc} \\ \hline\hline
&&\multicolumn{3}{l|}{JET FRAME SYNCHROTRON
PARAMETERS}&\multicolumn{2}{c}{HALFLIFE OBSERVED} \\
&&&&&\multicolumn{2}{c}{AT EARTH} \\
Source&knot&$\gamma$~range&$\gamma$~range&B$^{\prime}$(eq)&$\nu_s$ (max)&$\tau$[$\nu_s$(max)] \\
&&(synchrotron)&(0.3-8 keV)&($\mu$G)&(Hz)&(years) \\ \hline
&&&&&& \\
M87&HST-1&30-9.4E5&12- 61	&7&1E15&4.0 \\
        &D&22-8.7E5&20-107	&10&1E15&7.7 \\
        &A&29-9.1E5&34-174	&15&1E15&38. \\
        &B&31-9.8E5&56-290	&22&1E15&168. \\
&&&&&& \\      
3C 120 &25"& 77-2.4E4&17-89& 1.1&1E11& 153. \\
&&&&&& \\      
3C 273 &A1 & 40-1.3E6&27-139	& 6.9&1E15&  7.6 \\
       &B1  & 45-1.4E6& 34-174	& 6.7&1E15& 13.8 \\
       &D/H&35-1.1E6& 67-348	&22.&1E15& 132. \\
&&&&&& \\
PKS0637&wk7.8 &45-1.4E6& 68-351	&20.&1E15& 39. \\
&&&&&& \\
PKS1127&inner&60-1.9E4&130- 669&28.&1E11&8348. \\
       &rect.&147-4.6E4&232-1200& 8.3&1E11&20300. \\
       &A &200-6.3E4&182- 940& 3.5&1E11&7195. \\
       &B &134-4.2E4&306-1580&13.2&1E11&49500. \\
       &C &134-4.2E4&449-2318&19.&1E11&138700. \\ \hline
\end{tabular}

\noindent
NOTES                                                                    

\begin{description}
\item LENGTH:  'DfC' is distance from the core in arcsec, projected
kpc,\newline and de-projected kpc (= projected/sin$\theta$) for no bending.

\item The gamma range for the radio (synchrotron spectrum) corresponds
to the frequency range of 1~MHz to the value of $\nu_s$(max) given in
a later column.  The synchrotron spectrum is assumed to extend down to
1~MHz in order to include the low energy electrons required for the
beaming model.

\item $\nu_s$(max) is the highest reliable synchrotron frequency:
radio (E11 Hz) or optical (E15 Hz).

\item $\tau$ is the halflife for electrons that produce $\nu_s$(max).

\item R$^{\prime}$ is the ratio of IC to synchrotron losses in the jet frame.
\end{description}

\end{table}

\end{document}